%% file: tilings.tex
\documentclass[aps,pre,twocolumn,superscriptaddress,nofootinbib]{revtex4-2}
\usepackage{graphicx}
\usepackage{amsmath}
\usepackage{bm}
\usepackage{siunitx}

\begin{document}

\title{Configuration space partitioning in tilings of a bounded region of
the plane}

\author{Eduardo J. Aguilar}
\affiliation{Instituto de Ciência e Tecnologia,
Universidade Federal de Alfenas,
Rod.\ José Aurélio Vilela, 11999, 37715-400 Poços de Caldas - MG,
Brazil}

\author{Valmir C. Barbosa}
\email[]{valmir@cos.ufrj.br}
\affiliation{Programa de Engenharia de Sistemas e Computação, COPPE,
Universidade Federal do Rio de Janeiro,
Centro de Tecnologia, Sala H-319, 21941-914 Rio de Janeiro - RJ, Brazil}

\author{Raul Donangelo}
\affiliation{Instituto de Física, Facultad de Ingeniería,
Universidad de la República,
Julio Herrera y Reissig 565, 11.300 Montevideo, Uruguay}
\affiliation{Instituto de Física,
Universidade Federal do Rio de Janeiro,
Centro de Tecnologia, Bloco A, 21941-909 Rio de Janeiro - RJ, Brazil}

\author{Sergio R. Souza}
\affiliation{Instituto de Física,
Universidade Federal do Rio de Janeiro,
Centro de Tecnologia, Bloco A, 21941-909 Rio de Janeiro - RJ, Brazil}
\affiliation{Departamento de Engenharia Nuclear,
Universidade Federal de Minas Gerais,
Av.\ Antônio Carlos, 6627, 31270-901 Belo Horizonte - MG, Brazil}
\affiliation{Departamento de Física, ICEx,
Universidade Federal Fluminense,
R.\ Desembargador Ellis Hermydio Figueira, Aterrado,
27213-145 Volta Redonda - RJ, Brazil}

\begin{abstract}
Given a finite collection of two-dimensional tile types, the field of study
concerned with covering the plane with tiles of these types exclusively has a
long history, having enjoyed great prominence in the last six to seven decades,
not only as a topic of recreational mathematics but mainly as a topic of great
scientific interest. Much of this interest has revolved around fundamental
geometrical problems such as minimizing the variety of tile types to be used,
and also around important applications in areas such as crystallography as well
as others concerned with various atomic- and molecular-scale phenomena. All
these applications are of course confined to finite spatial regions, but in many
cases they refer back directly to progress in tiling the whole, unbounded plane.
Tilings of bounded regions of the plane have also been actively studied, but in
general the additional complications imposed by the boundary conditions tend to
constrain progress to mostly indirect results, such as recurrence relations, for
example. Here we study the tiling of rectangular regions of the plane by
rectangular tiles. The tile types we use are squares, dominoes, and straight
tetraminoes. For this set of tile types, not even recurrence relations seem to
be available. Our approach is to seek to characterize this complex system
through some fundamental physical quantities. We do this on two parallel tracks,
one fully analytical for what seems to be the most complex special case still
amenable to such approach, the other based on the Wang-Landau method for
state-density estimation. Given a simple energy function based solely on tile
contacts, we have found either approach to lead to illuminating depictions of
entropy, temperature, and above all partitions of the configuration space. The
notion of a configuration, in this context, refers to how many tiles of each
type are used. We have found that certain partitions help bind together
different aspects of the system in question and conjecture that future
applications will benefit from the possibilities they afford.
\end{abstract}

\maketitle

\section{Introduction}
\label{sec:int}

A tiling of the plane, given a finite collection $C$ of tile types, is a
covering that allows for no superposition of tiles and no gaps between them
while employing tiles of types in $C$ and no others. Depending on the tile types
that constitute it, $C$ is said to be periodic, non-periodic, or aperiodic. It
is periodic if its tile types only admit patterns that repeat themselves
indefinitely, which is the case, for example, of the regular hexagon as the only
tile type in $C$. If $C$ admits not only such repetitiveness but also indefinite
pattern diversification, for example when it only comprises the equilateral
triangle or the square, then it is called non-periodic. $C$ is called aperiodic
when indefinite diversification is the only possibility.

The first aperiodic tile set to be discovered dates from 1966 \cite{b66} and
uses over twenty thousand types of the edge-colored square tiles known as Wang
tiles. Tiling the plane with Wang tiles had been devised in the context of
studying the decidability of decision problems \cite{w61}, and as such required
tiling rules beyond the prohibition of superpositions or gaps. The resulting
aperiodicity quickly sparked an interest for finding smaller tile-type sets,
which within a few years led to the discovery of an aperiodic set with only six
types \cite{r71}. The well-known Penrose aperiodic sets were then soon
discovered \cite{p74,p78,g97}, first with six edge-marked types (three varieties
of the regular pentagon and three other shapes to fill gaps, thereby completing
work that Kepler had undertaken in the early 17th century), then with either one
dart and one kite, as the tiles became known, or two rhombi. The first aperiodic
monotile (a so-called ``einstein,'' a single tile type with which the plane can
be tiled without ever incurring periodicity) was discovered only very recently
\cite{smkg23a}. This monotile is one of the $873$ eight-kite polykites
\cite{w-pk} and has been named the hat. Notably, a polykite's basis kite is not
the same as the Penrose kite. The discovery of the hat was soon followed by that
of specters, monotiles closely related to the hat but having the property of
being chiral, i.e., of tiling the plane without ever being reflected
\cite{smkg23b}.

The two-rhombus Penrose tile set, both as proposed and in a generalization to
three dimensions, has had enduring impact in important fields, particularly in
crystallography, where the patterns it generates were found to suggest
structural ordering outside the classical approach (see \cite{m82} and
references therein). This connection foreshadowed the discovery of quasicrystals
that soon followed \cite{sbgc84} and has since continued to influence the field
\cite{zgbcult}. Other applications of the Penrose tile sets include modeling
jammed solids \cite{sl14} and the study of graph-theoretic properties of the
classical dimer model \cite{fsp20}. Beyond the direct applicability of specific
aperiodic tile sets, the very notion of their aperiodicity has had far-reaching
influence, e.g., in interpreting the results of self-assembled crystal
structures from molecular building blocks \cite{pd23} and in demonstrating the
use of ``algorithmic'' self-assembly of DNA strands into cellular automata
(specifically, one based on Wolfram's elementary rule 90, the XOR rule
\cite{w83}; see \cite{rpw04} and references therein). In fact, self-assembled
systems are now part of cutting-edge research in various fields, as in materials
science \cite{klclhjh22} and DNA-based computing
\cite{wdmhzyw19,dfglllnssyz21,xcs22,kljlklkck23}.

Of course, all these applications that ultimately refer back to tilings of the
plane are actually confined to finite regions, where the periodicity of the
tile-type collection $C$ ceases to have meaning, as do the issues regarding
decidability that helped spark the whole field decades ago. Instead, given
finiteness, the focus shifts to counting the tilings that $C$ admits, with
worries concerning undecidability giving way to the concrete possibility of
computational intractability. In fact, even though finding tilings of a finite
region of the plane can be achieved by solving a binary integer programming
problem \cite{gb20}, in general this problem is computationally intractable in
the sense of NP-hardness \cite{c71,k72}.

Here we consider the tiling of rectangular regions when $C$ contains rectangles
exclusively. We study the remarkably complex system that results, even for a
close to minimal $C$, as the great variety of possible tilings is taken into
account. Using an energy function that depends only on inter-tile contact, we
study entropy, temperature, and most importantly some key partitions of what we
call the system's configuration space. We use some of these terms in analogy to
their use in statistical physics, but develop specific meanings for the context
at hand. Doing this has been customary in more than one area, perhaps starting
with the landmark introduction of simulated annealing \cite{kgv83} and Hopfield
neural networks \cite{h82} in the early 1980s, the latter now generalized to
include Markov random fields \cite{ks80} and their many variations for use in
artificial intelligence, all sharing the Boltzmann-Gibbs multivariate
distribution as underlying statistical model \cite{p88,h90,kf09}; and more
recently, e.g., with the introduction of a variety of polygon-based models for
the study of structure and dynamics in biological tissues
\cite{fraej07,ags17,bhws17}. In the context of tiling a finite region of the
plane, focusing on how many tiles of each type in $C$ are used provides a means
to highlight specific interactions between tile structure, energy, and entropy.
We do succeed in describing a special case fully analytically, but in the
general case we resort to the Wang-Landau method to estimate state density
\cite{wl01a,wl01b}, with states now understood as tilings, before analysis can
be carried out.

We proceed as follows. We briefly review the current knowledge of tiling
rectangles by rectangles in Sec.~\ref{sec:til}, where we also introduce the $C$
and energy function we use. In Sec.~\ref{sec:spe} we analyze the special case we
mentioned and in Sec.~\ref{sec:gen} we tackle the general case. We conclude in
Sec.~\ref{sec:con}.

\section{Tilings of a rectangular board with rectangles}
\label{sec:til}

In order to study the statistical properties of tilings with rectangular tiles
of an $m\times n$ rectangular board (a board with $mn$ $1\times 1$ cells), the
central entity to be considered is the number $k_{mn}$ of possible distinct
tilings. Obtaining a closed-form expression for $k_{mn}$ depends not only on the
values of $m$ and $n$ but also on which tile types are to be used. Notably,
already for arbitrary $m,n\ge 2$ it seems that such an expression is known only
for the case in which dominoes ($1\times2$ or $2\times 1$ tiles) are the only
tiles used. In this case, each tiling requires $mn/2$ dominoes and $k_{mn}$ is
given by the surprising formula
\begin{equation}
k_{mn}=
\prod_{i=1}^{\lceil\frac{m}{2}\rceil}
\prod_{j=1}^{\lceil\frac{n}{2}\rceil}
\left(
4\cos^2\frac{\pi i}{m+1}+4\cos^2\frac{\pi j}{n+1}
\right),
\end{equation}
which as required is nonzero if and only if $mn$ is even
\cite{k61,tf61}.\footnote{If both $m$ and $n$ are odd, then
$\lceil m/2\rceil=(m+1)/2$ and $\lceil n/2\rceil=(n+1)/2$, which leads to a zero
factor.} If more tile types are to be used, then recurrence relations and
generating functions can still be obtained, but only in a limited manner. In
fact, fixing $m=2$ and leaving $n$ unconstrained while squares ($1\times 1$
tiles) and dominoes are the allowed tile types seems to be as far as one can go
(see, e.g., \cite{ks09}).

In this study we consider tilings with squares, dominoes, and straight
tetraminoes ($1\times 4$ or $4\times 1$ tiles, henceforth simply tetraminoes)
exclusively, though methodologically it is in principle possible to generalize
to a greater variety of tile types. We denote by $n_1$ the number of squares to
be used in a tiling, by $n_2$ the number of dominoes, and by $n_4$ the number of
tetraminoes. We refer to a joint assignment of values to $n_1,n_2,n_4$ admitting
at least one tiling of the board as a configuration of the system. Any
configuration implies $mn=n_1+2n_2+4n_4$, though this condition is in general
not sufficient for the assignment in question to qualify as a configuration (but
see the special case in Sec.~\ref{sec:spe}, where sufficiency clearly holds).

Given a configuration of the system, and considering any tiling it admits, let
$u$ and $v$ be any two of the tiles used. We denote by $[u\vert v]$ the
tile-perimeter length that is common to $u$ and $v$. It follows that
$[u\vert v]>0$ if and only if $u$ and $v$ are adjacent to each other in the
tiling in question. Additionally, the sum of $[u\vert v]$ over all $u,v$ pairs
is conserved over all possible tilings for the same configuration, since in any
such tiling every tile contributes half its perimeter to the sum, discounting
those tile edges that coincide with the board's own perimeter and therefore
contribute nothing. This sum quantifies all inter-tile contacts and is here used
as the system's energy function. That is,
\begin{align}
E(n_1,n_2)
&=\sum_{\genfrac{}{}{0pt}{}{u,v\in I(n_1,n_2)}{u\neq v}}[u\vert v] \\
&=2n_1+3n_2+5n_4-m-n,
\label{h}
\end{align}
where $I$ is the set comprising distinguishable versions of all $n_1+n_2+n_4$
tiles (so that $I$ is in fact a set). In the above expression we write both $E$
and $I$ as functions of only $n_1$ and $n_2$ to highlight the simple fact that,
given $m$ and $n$, one of $n_1,n_2,n_4$ is necessarily a function of the other
two. We have chosen $n_4$ for this role, so its value is to be determined as
\begin{equation}
n_4=\frac{mn-n_1-2n_2}{4}.
\label{n4}
\end{equation}

It is often possible for energy levels $\varepsilon$ to exist such that
$E(n_1,n_2)=\varepsilon$ for more than one configuration of the system. For the
purpose of discussing entropy and temperature, we handle such ``degeneracy''
both by focusing on each of the configurations involved independently of the
others and by taking them into account together. Doing this allows for distinct
perspectives from which to analyze the system.

\section{A special case}
\label{sec:spe}

The most complex systems for which analytical treatment is possible in this
three tile-type scenario seem to be those for which $m=1$, that is, those whose
board is $1\times n$. In this case, we have
\begin{equation}
n_4=\frac{n-n_1-2n_2}{4}
\end{equation}
and
\begin{equation}
E(n_1,n_2)=n_1+n_2+n_4-1.
\label{Em=1}
\end{equation}
Moreover, given a configuration, the number of tilings it admits, now expressed
as a function of only $n_1,n_2$ as well, is 
\begin{equation}
k_{1n}(n_1,n_2)=(n_1,n_2,n_4)!=\frac{(n_1+n_2+n_4)!}{n_1!\,n_2!\,n_4!},
\end{equation}
which is the multinomial coefficient for $n_1,n_2,n_4$. To continue, we first
rewrite $k_{1n}$ as
\begin{equation}
k_{1n}(n_1,n_2)=
\frac{\Gamma(n_1+n_2+n_4+1)}{\Gamma(n_1+1)\,\Gamma(n_2+1)\,\Gamma(n_4+1)}
\label{k1nG}
\end{equation}
and whenever needed substitute the reals $x,y$ for the integers $n_1,n_2$,
respectively, so that differentiation can be carried out properly. We also note
that
\begin{equation}
\frac{\partial}{\partial w}\ln\Gamma(w)=\frac{\Gamma'(w)}{\Gamma(w)}=\psi_0(w),
\end{equation}
where $\psi_0$ is the digamma function.

For $k>0$ an integer, we have $\psi_0(k)=-\gamma$ if $k=1$,
$\psi_0(k)=\sum_{\ell=1}^{k-1}\ell^{-1}-\gamma$ if $k>1$, where
$\gamma\approx 0.5772$ is the Euler constant. It follows, e.g., that
$\Gamma'(w)\approx 0.4228,1.8456,7.5366,36.1464$ for $w=2,3,4,5$, while an
approximation of what could pass for the ``derivative'' of the factorial
function at the integer $w-1$, given by
\begin{equation}
\frac{\Delta\Gamma(w)}{2}=\frac{\Gamma(w+1)-\Gamma(w-1)}{2},
\end{equation}
yields $2^{-1}\Delta\Gamma(w)=0.5,2.5,11,57$ for the same values of $w$. In
addition, $\Gamma'(1)$ is unique in that it is negative. Thus, even though we do
in the sequel use $\Gamma'(w)$ as a measure of the local variability of the
factorial function at $w-1$, some inconsistency is to be expected. We return to
this in our analysis in Sec.~\ref{ssec:dg-1xn}.

For later reference, we note further that, by Eq.~(\ref{Em=1}) and letting
$a=n\bmod 4$, the possible values of $E$ range from a minimum that uses as many
tetraminoes as possible ($n_4=\lfloor n/4\rfloor$), and also as few dominoes
($n_2=\lfloor a/2\rfloor$) and squares ($n_1=a\bmod 2$) as possible, to a
maximum that only uses squares ($n_1=n,n_2=n_4=0$). This yields
\begin{align}
E_\mathrm{min}^{1\times n} &=
a\bmod 2+\lfloor a/2\rfloor+\lfloor n/4\rfloor-1 \\
&= \lceil a/2\rceil+\lfloor n/4\rfloor-1
\end{align}
and
\begin{equation}
E_\mathrm{max}^{1\times n}=n-1.
\end{equation}

\subsection{The nondegenerate case}

\begin{figure}[t]
\includegraphics[scale=0.68]{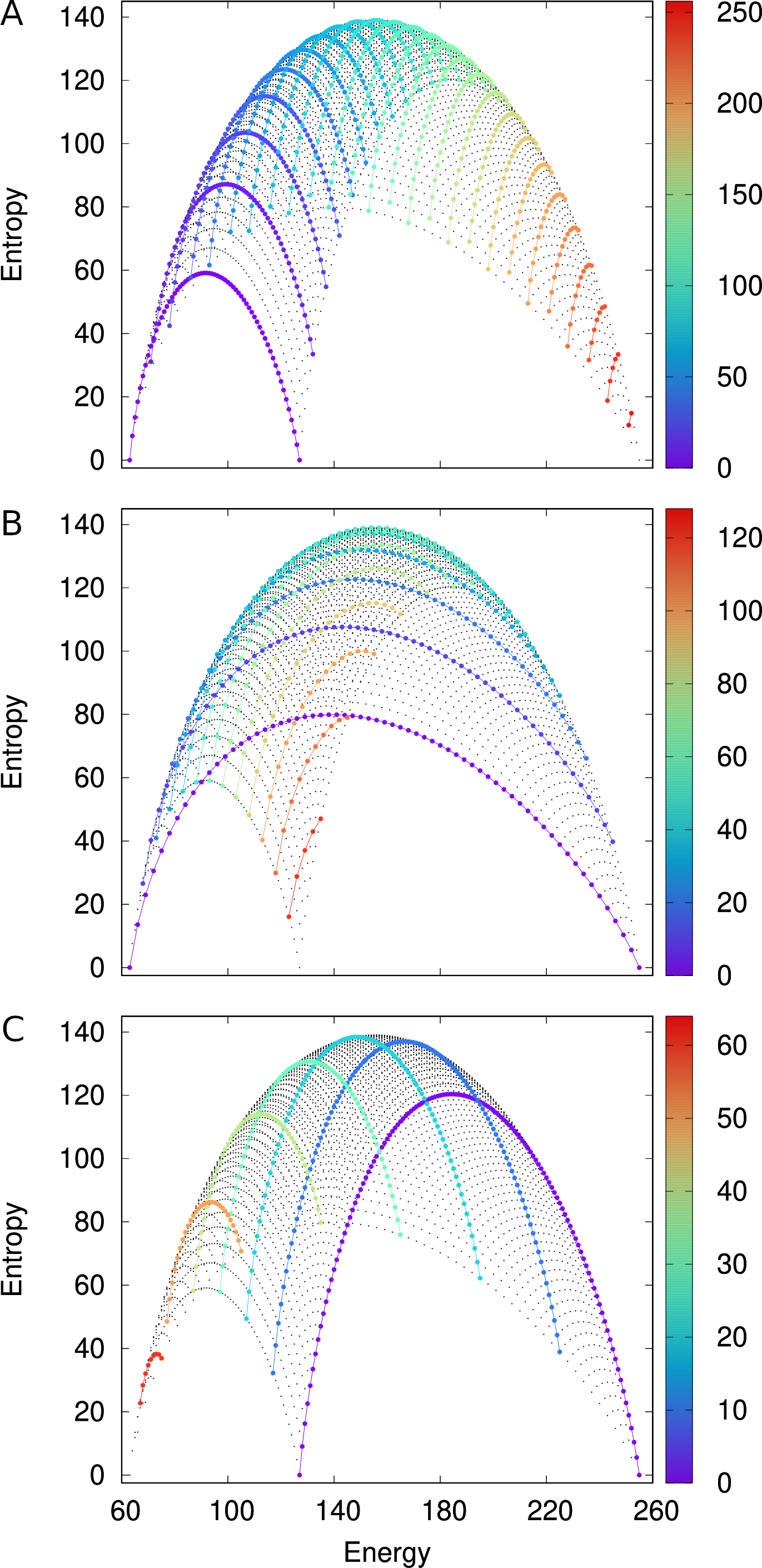}
\caption{Partitioning the configuration space for $m=1$, $n=256$. Each panel
shows a different partition of all $4\,225$ configurations (dots), each
represented by its energy $E$ and entropy $S_\mathrm{nd}$. A set in a partition
is shown as a color-coded thread of enlarged dots and corresponds to a fixed
value of $n_1$ (A), $n_2$ (B), or $n_4$ (C). A thread's color is picked from the
panel's color bar on the right for that fixed value. To avoid cluttering, only
sets corresponding to multiple-of-$10$ values of $n_1$ are shown in panel A, and
likewise for $n_2$ in panel B and $n_4$ in panel C.
$E_\mathrm{min}^{1\times 256}=63$, $E_\mathrm{max}^{1\times 256}=255$.}
\label{ndg-1xn}
\end{figure}

Analyzing the system from a configuration-centric perspective (i.e., by
essentially ignoring the possible degeneracy of certain energy levels) amounts
to regarding both entropy and temperature as functions of $n_1,n_2$. That is,
\begin{equation}
S_\mathrm{nd}(n_1,n_2)=\ln k_{1n}(n_1,n_2)
\end{equation}
and, using Eq.~(\ref{k1nG}) with $\Gamma'(w)=\Gamma(w)\psi_0(w)$,
\begin{align}
T^{-1}_\mathrm{nd}(n_1,n_2)
&=
\left(
\frac{\partial S_\mathrm{nd}}{\partial x}\frac{\partial x}{\partial E}+
\frac{\partial S_\mathrm{nd}}{\partial y}\frac{\partial y}{\partial E}
\right)
\biggr\rvert_{\genfrac{}{}{0pt}{}{x=n_1}{y=n_2}} \\
&=
U(n_1,n_2),
\end{align}
where
\begin{align}
U(n_1,n_2)
={}&2\psi_0(n_1+n_2+n_4+1)-{} \\
&\frac{4}{3}\,\psi_0(n_1+1)-2\psi_0(n_2+1)+\frac{4}{3}\,\psi_0(n_4+1). \nonumber
\end{align}

\begin{table*}[t]
\caption{Details of the set of configurations for $n=25$ and $n_4=1$.}
\label{tab-ndg-1xn}
\centering
\input{table1}
\end{table*}

Partitioning the configuration space can be achieved in this case by fixing the
value of, say, $n_1$ and observing the configurations that result as $n_2$ and
$n_4$ are varied. Taking $n=256$, for example, results in the $4\,225$
configurations shown as background dots in the $E\times S_\mathrm{nd}$ plots of
Fig.~\ref{ndg-1xn}. Each of panels A, B, C in the figure corresponds to a
different partition of the configuration space, showing some of the sets that
result from assigning fixed values to $n_1$, $n_2$, $n_4$, respectively. Each
set can be seen to be generally characterized by a ``smooth'' succession of
points along which the values of energy $E$ increase while the values of entropy
$S_\mathrm{sd}$ first rise then decline. For a more manageable value of $n$
($n=25$), we give all details of the $n_4=1$ set in Table~\ref{tab-ndg-1xn},
where $k_{1n}^{n_4=1}=k_{1n}\bigr\rvert_{n_4=1}$, including an illustration of
one of the $k_{1n}^{n_4=1}$ tilings for each pair $n_1,n_2$ of values. Clearly,
the set in question is characterized by an initial preponderance of dominoes
that, along the sequence of increasing values of $E$, eventually turns into a
preponderance of squares. 

\begin{figure}[t]
\includegraphics[scale=0.68]{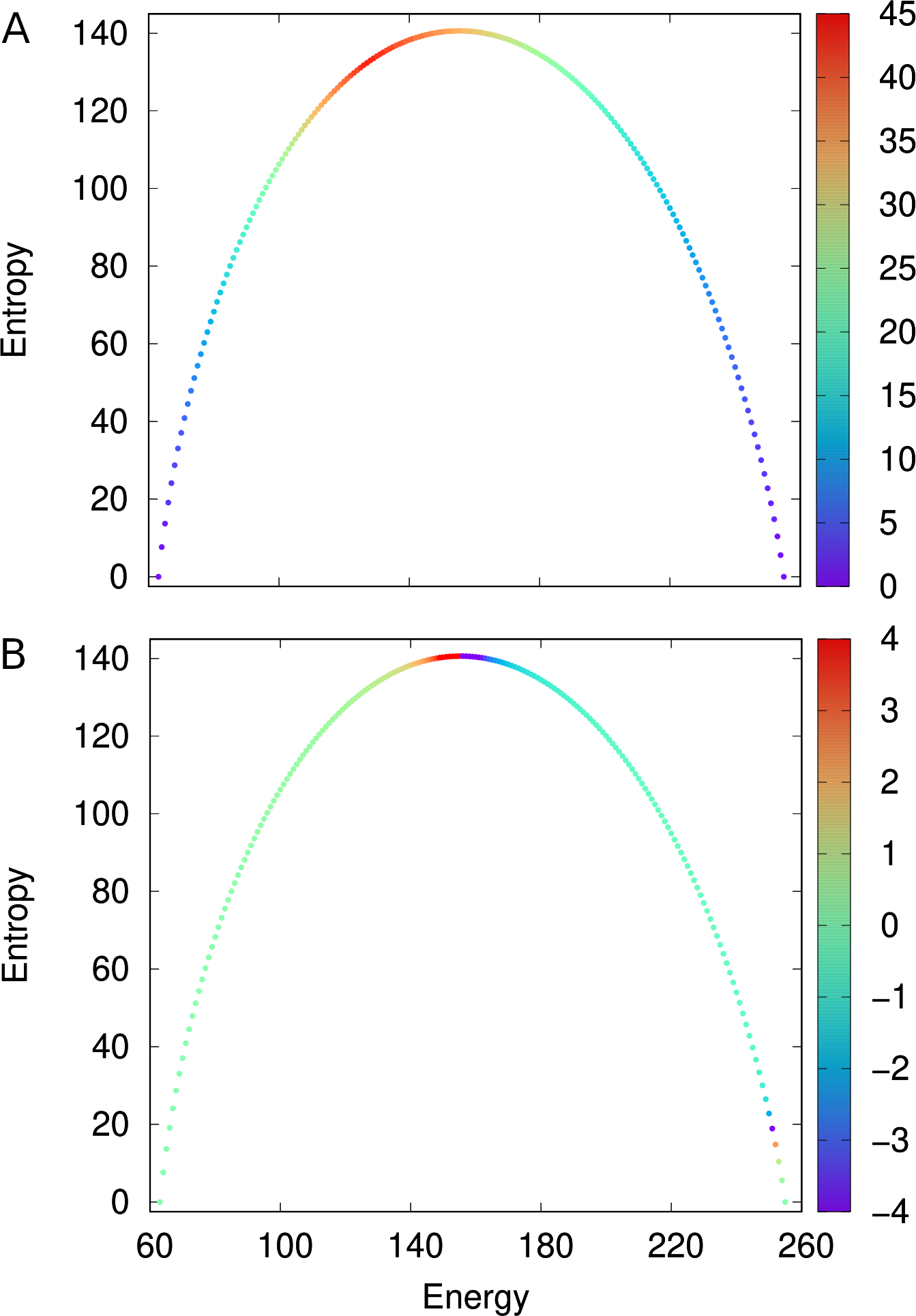}
\caption{Entropy $S_\mathrm{d}$ as a function of energy $E$ for $m=1$, $n=256$.
Dots are color-coded according to each panel's color bar on the right to
indicate the number of configurations (A) and the temperature $T_\mathrm{d}$ (B)
for each energy level. For readability, the color bar in (B) leaves out $7$
points for which $T_\mathrm{d}<-4$ and $6$ others for which $T_\mathrm{d}>4$.
These points get colored with the color corresponding to $-4$ or to $4$,
respectively.
$E_\mathrm{min}^{1\times 256}=63$, $E_\mathrm{max}^{1\times 256}=255$.}
\label{deg-1xn}
\end{figure}

\subsection{The degenerate case}
\label{ssec:dg-1xn}

An alternative analysis strategy is to recognize the existence of degenerate
energy levels and take it fully into account. For each value $\varepsilon$ in
the interval from $E_\mathrm{min}^{1\times n}$ to $E_\mathrm{max}^{1\times n}$,
the system's entropy and temperature are functions of $\varepsilon$. Entropy is
given by
\begin{equation}
S_\mathrm{d}(\varepsilon)=\ln K(\varepsilon),
\end{equation}
where
\begin{equation}
K(\varepsilon)=
\sum_{\genfrac{}{}{0pt}{}{n_1,n_2}{E(n_1,n_2)=\varepsilon}}k_{1n}(n_1,n_2),
\label{Kh}
\end{equation}
and temperature is such that
\begin{align}
T^{-1}_\mathrm{d}(\varepsilon)
&=
\frac{\partial S_\mathrm{d}}{\partial x}\frac{\partial x}{\partial E}+
\frac{\partial S_\mathrm{d}}{\partial y}\frac{\partial y}{\partial E} \\
&=
\frac{1}{K(\varepsilon)}
\sum_{\genfrac{}{}{0pt}{}{n_1,n_2}{E(n_1,n_2)=\varepsilon}}
\left(
\frac{\partial k_{1n}}{\partial x}\frac{\partial x}{\partial E}+
\frac{\partial k_{1n}}{\partial y}\frac{\partial y}{\partial E}
\right)
\biggr\rvert_{\genfrac{}{}{0pt}{}{x=n_1}{y=n_2}} \\
&=
\sum_{\genfrac{}{}{0pt}{}{n_1,n_2}{E(n_1,n_2)=\varepsilon}}
\frac{k_{1n}(n_1,n_2)}{K(\varepsilon)}\,U(n_1,n_2).
\label{tdgm1}
\end{align}
That is, $T^{-1}_\mathrm{d}(\varepsilon)$ is a convex combination of
$U(n_1,n_2)$ for those configurations for which $E(n_1,n_2)=\varepsilon$.

Thus, as far as representing configurations by their $E$ and $S_\mathrm{d}$
values is concerned, all configurations for which the value of $E$ is the same
become conjoined in an $E\times S_\mathrm{d}$ plot. This is illustrated in the
panels of Fig.~\ref{deg-1xn} for $n=256$. As expected, this figure's panel A
reveals a greater variety of configurations contributing to $S_\mathrm{d}$ near
the midrange values of $E$. As for panel B, overall we also see $T_\mathrm{d}$
behave as expected, that is, slightly above zero and increasing before
$S_\mathrm{d}$ peaks as $E$ grows, then abruptly negative and increasing
toward slightly below zero. However, a closer examination reveals a sudden flip
back to positive temperatures for the highest four values of $E$: for $E=252$
and only one of the two contributing configurations ($n_1=252,n_2=0,n_4=1$); for
$E=253$ and the only contributing configuration ($n_1=252,n_2=2,n_4=0$); for
$E=254$ and the only contributing configuration ($n_1=254,n_2=1,n_4=0$); and for
$E=255$ and the only contributing configuration ($n_1=256,n_2=n_4=0$).

\begin{figure}[b]
\includegraphics[scale=0.68]{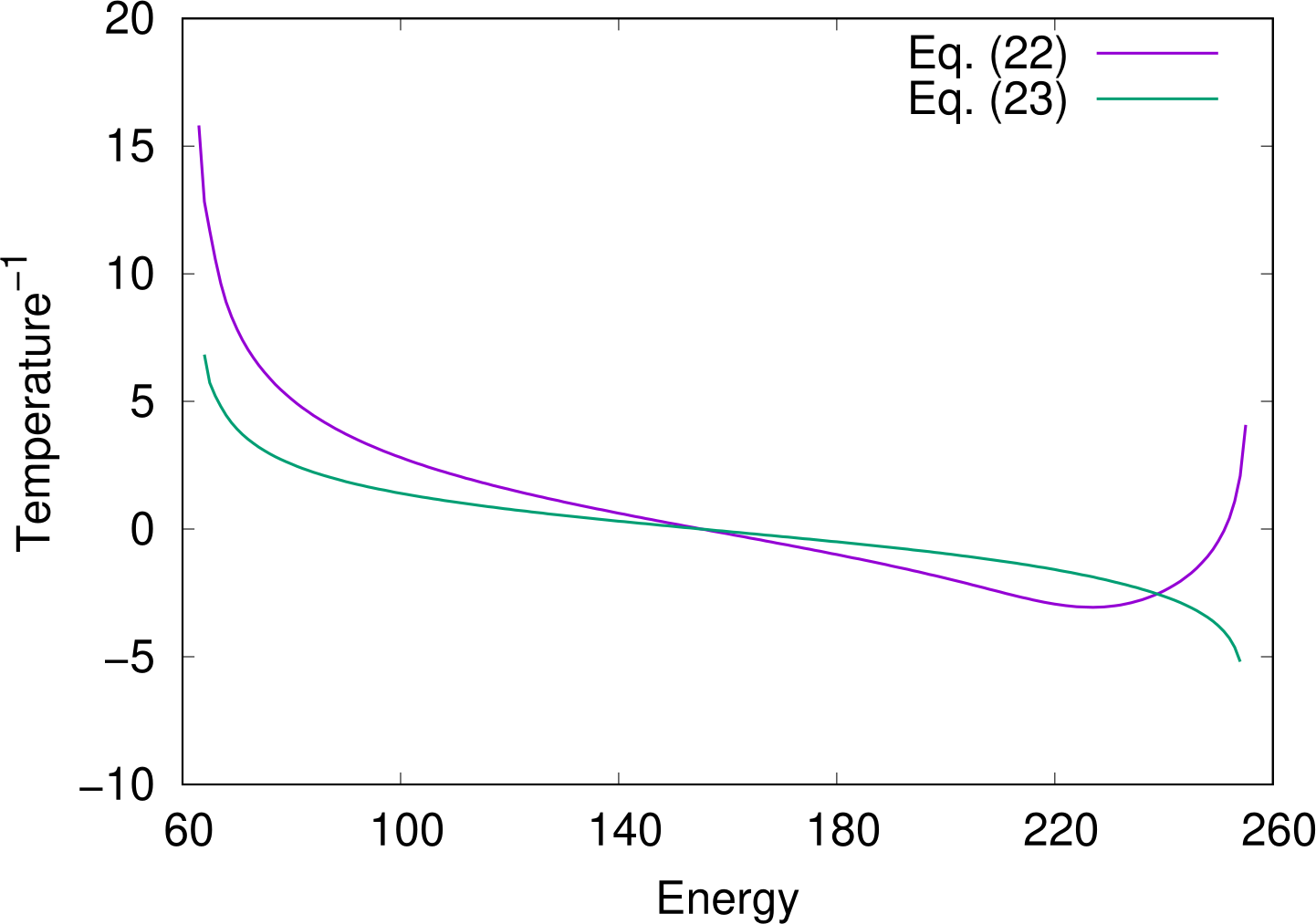}
\caption{$T_\mathrm{d}^{-1}$ as a function of energy $E$ according to
Eqs.~(\ref{tdgm1}) and~(\ref{tdgm1d}).}
\label{deg-1xn-t}
\end{figure}

This can be further explored as in Fig.~\ref{deg-1xn-t}, which illustrates the
behavior of $T_\mathrm{d}^{-1}$ according to both Eq.~(\ref{tdgm1}) and
\begin{equation}
T_\mathrm{d}^{-1}(\varepsilon)\approx
\frac{S_\mathrm{d}(\varepsilon+1)-S_\mathrm{d}(\varepsilon-1)}{2}.
\label{tdgm1d}
\end{equation}
The sudden turn to positive temperatures described above is clearly visible, as
is the overall discrepancy between the two curves. We attribute these
differences to the problems that are inherent to using $\Gamma'$ to assess the
local variability of the factorial function, as discussed in the introduction to
Sec.~\ref{sec:spe}. In spite of these difficulties, for most of the energy
spectrum Eq.~(\ref{tdgm1}) provides a reasonable representation of the actual
quantity. It is also significant that the use of $\Gamma'$ is the only available
analytical technique for temperature assessment in the case at hand.

\subsection{Remarks on computational tractability}
\label{ssec:ct}

\begin{figure}[t]
\includegraphics[scale=0.68]{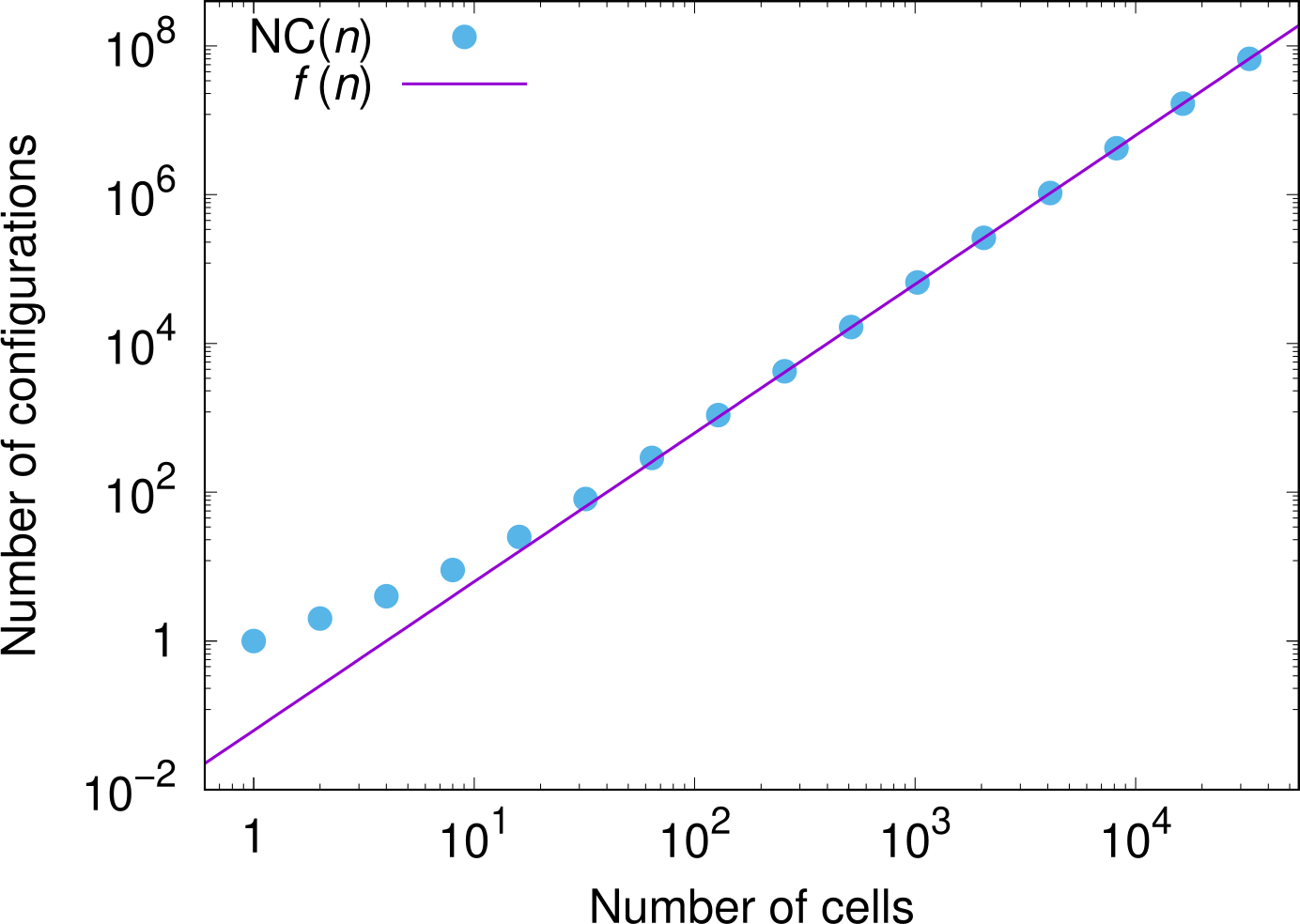}
\caption{Total number of configurations, for $m=1$, as the function
$\mathrm{NC}(n)$ of the number of cells $n$. The power law that fits the points
asymptotically is $f(n)=4^{-2}n^2$. This comes from
$\mathrm{NC}(n)=
\sum_{n_1=0}^n\sum_{n_2=0}^{\lfloor(n-n_1)/2\rfloor}
[(n-n_1-2n_2)\bmod 4=0]\approx
\sum_{n_1=0}^n\sum_{n_2=0}^{\lfloor(n-n_1)/2\rfloor}
4^{-1}$,
which equals $4^{-2}(n+1)(n+3)$ if $n$ is odd, $4^{-2}(n+2)^2$ if $n$ is even.
Either of these expressions tends asymptotically to $f(n)$. Counting each
configuration as only $4^{-1}$ targets the many situations where the condition
$(n-n_1-2n_2)\bmod 4=0$ fails, causing three out of four configurations to be
invalid.}
\label{nconfigs-1xn}
\end{figure}

In terms of the computational difficulties involved, an important point to note
if we were to move beyond $n=256$ is that obtaining plots like the ones in 
Figs.~\ref{ndg-1xn} and~\ref{deg-1xn} would become increasingly harder. This is
so because those plots contemplate all possible configurations of the system,
which for $n=1,2,\ldots,2^{15}$ grows from $1$ to $67\,125\,249$, as shown in
Fig.~\ref{nconfigs-1xn}. These numbers are not particularly impressive, but
already for $n=2^{16}$ we found that $128$ GB of memory were insufficient for
the Mathematica 13 system to generate all configurations. (Regarding notation,
in the caption of Fig.~\ref{nconfigs-1xn}, and henceforth, we use the Iverson
bracket $[P]$ for $P$ a logical proposition. $[P]$ equals $1$ if $P$ is true,
$0$ if $P$ is false. This notation generalizes the Kronecker delta, since
$[i=j]=\delta_{ij}$.)

\begin{figure}[t]
\includegraphics[scale=0.68]{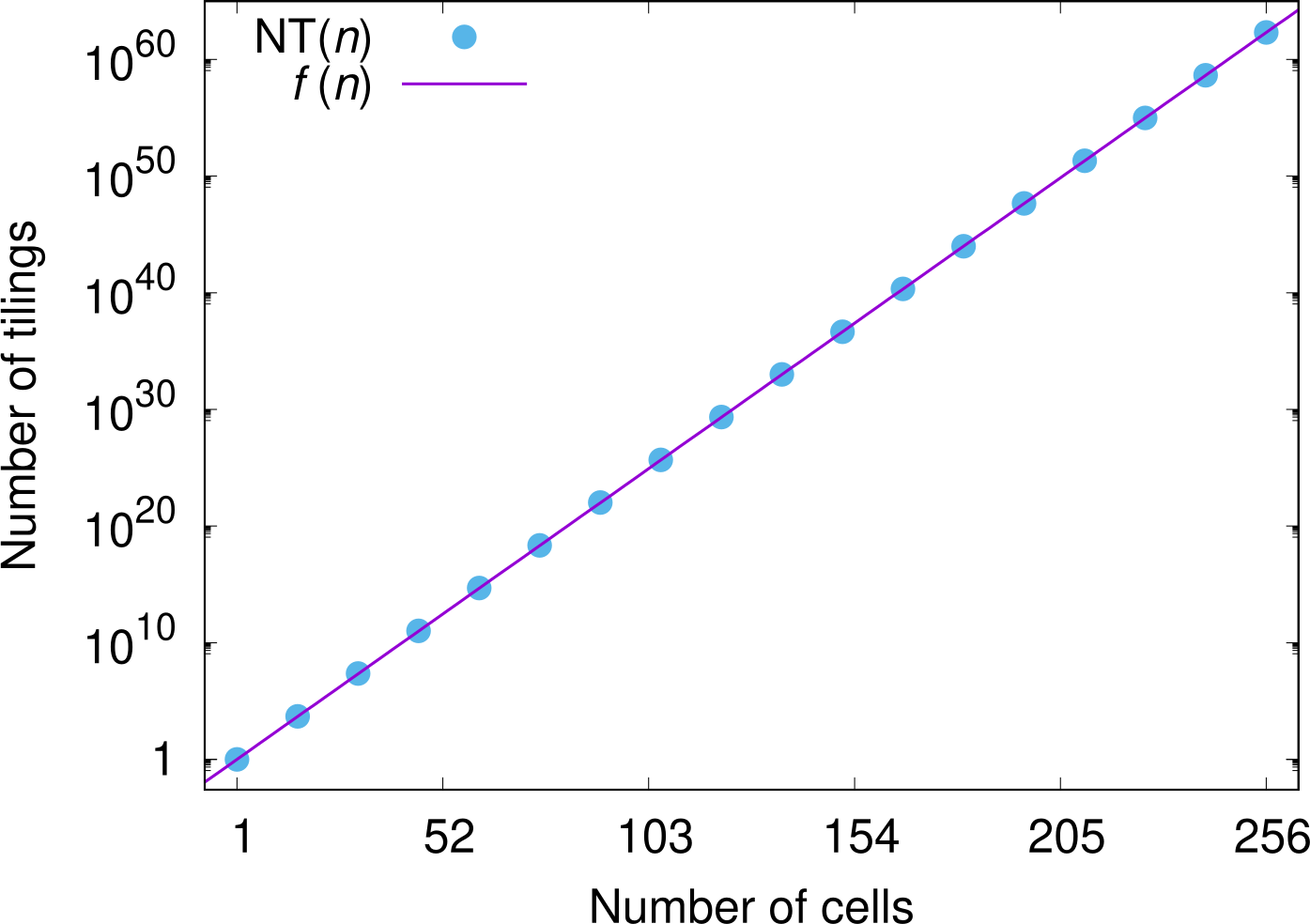}
\caption{Total number of tilings, for $m=1$, as the function
$\mathrm{NT}(n)=\sum_\varepsilon K(\varepsilon)$ of the number of cells $n$. The
exponential fitting the points is $f(n)=e^{0.562663(n-1)}$.}
\label{ntilings-1xn}
\end{figure}

On the other hand, it must be kept in mind that Figs.~\ref{ndg-1xn}
and~\ref{deg-1xn} could only be obtained due to the availability of the
closed-form expression for $k_{1n}$ given in Eq.~(\ref{k1nG}), which essentially
does away with the need to count the number of tilings admitted by each of the
configurations in order to calculate entropy. Totaling this number over all
configurations results in $\sum_\varepsilon K(\varepsilon)$, whose growth as a
function of $n$ is depicted in Fig.~\ref{ntilings-1xn}. Crucially, for $n=256$
the value of $\sum_\varepsilon K(\varepsilon)$ is already of the order of
$10^{63}$, which can be expected to be surpassed by many orders of magnitude as
we consider the $m>1$ cases even for boards with a similar number of cells.
Nothing like Eq.~(\ref{k1nG}) is known for $m>1$, so clearly generalizing the
special case of $m=1$ requires the ability to estimate entropy without counting
the number of tilings admitted by any given configuration.

\section{The general case}
\label{sec:gen}

For the general case we follow the same two strategies used in
Sec.~\ref{sec:spe}, i.e., we study both the nondegenerate case (by focusing on
individual configurations even though in general there can be several of them
for the same energy level) and the degenerate one. As previously, therefore, the
first strategy essentially sidelines the issue of energy-level degeneracy, thus
allowing some partitions of the configuration space to be highlighted as in
Fig.~\ref{ndg-1xn}. Our approach relies on sampling tilings randomly,
calculating the energy level for each sample, then adding some contribution to
the ongoing entropy estimate for that energy level.

\subsection{Entropy estimation}
\label{ssec:entrest}

Our approach to obtain entropy estimates is to use the Wang-Landau method
\cite{wl01a,wl01b}, which is a Monte Carlo Markov Chain (MCMC) method with
well-established convergence properties \cite{bp07,fjkls15} to estimate state
densities in statistical models. MCMC methods work by placing a walker at some
randomly chosen initial state and then having it hop from state to state
according to the chain's transition probabilities, recording information along
the way until some stopping criterion is met. The specific formulation we use is
derived from the Metropolis-Hastings method \cite{mrrtt53,h70}, in which the
transition probabilities are based on the chain's detailed-balance conditions
and therefore make the desired stationary distribution explicit. Upon
convergence to all detailed-balance conditions being satisfied, that is the
distribution that will be observed (see Sec.~\ref{ssec:succ}).

Given the values of $m$ and $n$, the state space in this study is the set of all
tilings of the $m\times n$ board by squares, dominoes, and tetraminoes. Each
tiling is relative to a configuration unequivocally specified by the values of
$n_1$ and $n_2$. A tiling's target probability in the desired stationary
distribution is proportional to $k_{mn}^{-1}(n_1,n_2)$, whose value is unknown
but can be estimated, once an estimate $\hat S(n_1,n_2)$ is available for the
corresponding configuration's entropy $S(n_1,n_2)$, as $e^{-\hat S(n_1,n_2)}$.
Thus, given two tilings $t$ and $t'\neq t$, of configurations $n_1,n_2$ and
$n'_1,n'_2$, respectively, the detailed-balance condition for them reads
\begin{equation}
e^{-\hat S(n_1,n_2)}g_{t\to t'}a_{t\to t'}=
e^{-\hat S(n'_1,n'_2)}g_{t'\to t}a_{t'\to t},
\label{dbc}
\end{equation}
where $g_{t\to t'}a_{t\to t'}$ is the transition probability from $t$ to $t'$,
broken down into the probability $g_{t\to t'}$ that tiling $t'$ is generated as
a possible successor to tiling $t$ and the probability $a_{t\to t'}$ of actually
accepting the succession.

Given $g_{t\to t'}$ and $g_{t'\to t}$, the acceptance probability of $t'$ given
$t$ is defined as
\begin{equation}
a_{t\to t'}=
\min\left\{1,\frac
{e^{-\hat S(n'_1,n'_2)}g_{t'\to t}}
{e^{-\hat S(n_1,n_2)}g_{t\to t'}}
\right\},
\label{accept}
\end{equation}
whence it follows that 
\begin{equation}
\frac{a_{t\to t'}}{a_{t'\to t}}=
\frac{e^{-\hat S(n'_1,n'_2)}g_{t'\to t}}
{e^{-\hat S(n_1,n_2)}g_{t\to t'}}.
\end{equation}
The definition in Eq.~(\ref{accept}), therefore, leads directly to the condition
in Eq.~(\ref{dbc}).

\subsection{Successor generation}
\label{ssec:succ}

We assume $g_{t\to t'}>0$ if and only if $t'$ can be obtained from $t$ either
through the split of one tile (a domino or a tetramino) into two tiles (two
squares or two dominoes, respectively) or through the merger of two tiles (two
adjacent squares or two longitudinally adjacent dominoes) into one single tile
(a domino or a tetramino, respectively). If $g_{t\to t'}>0$ does indeed hold,
then so does $g_{t'\to t}>0$. It follows that the Markov chain in question is
ergodic, that is, both aperiodic (since
$\sum_{t'}g_{t\to t'}a_{t\to t'}<1$, as there is always the possibility of
rejection \cite{addj03}, so whenever transitioning from $t$ it is possible to
remain at $t$) and irreducible (i.e., any state can be reached from any other).
Therefore, the chain has a unique stationary distribution, which as discussed in
Sec.~\ref{ssec:entrest} is proportional to $e^{-\hat S}$.

Let $n^t_\mathrm{s}$ be the number of dominoes or tetraminoes in tiling $t$, and
$n^t_\mathrm{m}$ the number of pairs of adjacent squares or pairs of
longitudinally adjacent dominoes in $t$. The generation of $t'$ from $t$ starts
with deciding which operation, the split of a domino into two squares or a
tetramino into two dominoes (with probability $p^t_\mathrm{s}$) or the merger of
two adjacent squares or two longitudinally adjacent dominoes (with probability
$p^t_\mathrm{m}=1-p^t_\mathrm{s}$), is to be applied. Probability
$p^t_\mathrm{s}$ is given by
\begin{equation}
p^t_\mathrm{s}=\frac
{[n^t_\mathrm{s}>0]}
{[n^t_\mathrm{s}>0]+[n^t_\mathrm{m}>0]},
\end{equation}
so $p^t_\mathrm{s}$ equals $0$, $2^{-1}$, or $1$, since
$n^t_\mathrm{s}+n^t_\mathrm{m}>0$ always holds. Once the decision of whether to
split or to merge has been made, the tile to be split or the pair of tiles to be
merged is chosen uniformly at random. We then have
\begin{equation}
g_{t\to t'}=\begin{cases}
(n^t_\mathrm{s})^{-1}
& \text{ if }t'\in T_\mathrm{s}^t\text{ and }n_\mathrm{m}^t=0, \\
(n^t_\mathrm{m})^{-1}
& \text{ if }t'\in T_\mathrm{m}^t\text{ and }n_\mathrm{s}^t=0, \\
(2n^t_\mathrm{s})^{-1}
& \text{ if }t'\in T_\mathrm{s}^t\text{ and }n_\mathrm{s}^tn_\mathrm{m}^t>0, \\
(2n^t_\mathrm{m})^{-1}
& \text{ if }t'\in T_\mathrm{m}^t\text{ and }n_\mathrm{s}^tn_\mathrm{m}^t>0, \\
0
& \text{ if }t'\notin T_\mathrm{s}^t\cup T_\mathrm{m}^t.
\end{cases}
\end{equation}
In this equation, $T_\mathrm{s}^t$ comprises the $n_\mathrm{s}^t$ tilings
obtainable from $t$ via a split and $T_\mathrm{m}^t$ comprises the
$n_\mathrm{m}^t$ tilings obtainable from $t$ via a merger.

\subsection{Implementation of the Wang-Landau method}

Our implementation of the Wang-Landau method follows the steps outlined next,
where $\mathrm{hist}_E$ and $\mathrm{hist}_{\hat S}$ are histograms to record
how many times each energy level is observed during the walker's traversal of
the Markov chain and this level's entropy estimate, respectively. A parameter
$f$ is used to control the entropy estimates. It is set to $1$ initially and is
decreased to half its current value at each reset of $\mathrm{hist}_E$.
Termination occurs when $f<10^{-10}$.

\begin{enumerate}
\item{}
$t\leftarrow t_0$, where $t_0$ is a randomly chosen tiling; \\
Let $\varepsilon_t$ be the energy level of $t$;
\item{}\label{reset}
$\mathrm{hist}_E(\varepsilon)\leftarrow 0$ for every applicable energy level
$\varepsilon$; \\
$\mathrm{hist}_{\hat S}(\varepsilon)\leftarrow 0$ for every applicable energy
level $\varepsilon$; \\
$f\leftarrow 1$; \\
$\mathrm{hist}_E(\varepsilon_t)\leftarrow 1$; \\
$\mathrm{hist}_{\hat S}(\varepsilon_t)\leftarrow f$; \\
Go to Step~\ref{continue};
\item{}\label{flat}
$\mathrm{hist}_E(\varepsilon)\leftarrow 0$ for every applicable energy level
$\varepsilon$;
\item{}\label{continue}
Generate a tentative successor $t'$ of $t$; \\
With probability $a_{t\to t'}$, do $t\leftarrow t'$; \\
Let $\varepsilon_t$ be the energy level of $t$; \\
If $\mathrm{hist}_E(\varepsilon_t)=0$, go to Step~\ref{reset}; \\
$\mathrm{hist}_E(\varepsilon_t)\leftarrow\mathrm{hist}_E(\varepsilon_t)+1$; \\
$\mathrm{hist}_{\hat S}(\varepsilon_t)\leftarrow
\mathrm{hist}_{\hat S}(\varepsilon_t)+f$; \\
If $\mathrm{hist}_E$ is flat, do $f\leftarrow 2^{-1}f$ and
go to Step~\ref{flat}; \\
If $f\ge 10^{-10}$, repeat Step~\ref{continue};
\end{enumerate}

The random walk starts at the randomly chosen tiling $t_0$ and is restarted
whenever an energy level not yet encountered is found. Both initially and when
a restart occurs, the two histograms $\mathrm{hist}_E$ and
$\mathrm{hist}_{\hat S}$ are reset in Step~\ref{reset}. Another opportunity for
a reset, albeit a partial one, occurs when $\mathrm{hist}_E$ becomes flat and is
then reset in Step~\ref{flat}. Flatness is detected in Step~\ref{continue}
whenever every energy level $\varepsilon$ encountered thus far has
$\mathrm{hist}_E(\varepsilon)$ no lower than $95\%$ of the histogram's average.
Except for the resets in Steps~\ref{reset} and~\ref{flat}, Step~\ref{continue}
keeps repeating until termination occurs. After this, for each energy level
$\varepsilon$ reached by the walker the entropy estimate is relativized to that
of energy level
$\varepsilon^*=\arg\min_{\varepsilon'}\mathrm{hist}_{\hat S}(\varepsilon')$, via
$\mathrm{hist}_{\hat S}(\varepsilon)\leftarrow
\mathrm{hist}_{\hat S}(\varepsilon)-\mathrm{hist}_{\hat S}(\varepsilon^*)$. In a
run that reaches every energy level, $\varepsilon^*$ corresponds to the
configuration $n_1=mn,n_2=0$, that is, $\varepsilon^*=2mn-m-n$ and
$k_{mn}(n_1,n_2)=1$. In this case, the $\mathrm{hist}_{\hat S}(\varepsilon^*)=0$
resulting from relativization is no longer an estimate but the exact value. Note
that this holds automatically in the $m=1$ case of Sec.~\ref{sec:spe}.

\subsection{Handling degeneracy}

As we normally do when handling the degenerate case, so too in the nondegenerate
case we would like to associate an entropy estimate to each energy level
directly. However, in general an energy level does not unequivocally determine a
system configuration, so we opt instead to slightly alter the counting of
squares and dominoes in a tiling. This is done by letting squares be counted in
units of weight $1+\delta_1$ and dominoes in units of weight $1+\delta_2$. The
values of $\delta_1,\delta_2$ must be sufficiently small to not affect the value
of $E$ in any meaningful way, while allowing any two distinct configurations
that would otherwise have the same value of $E$ to be told apart from each other
by the now slightly different values of $E$.

In practice, this amounts to substituting $n_1(1+\delta_1)$ for $n_1$ in
Eqs.~(\ref{h}) and~(\ref{n4}), and likewise $n_2(1+\delta_2)$ for $n_2$. From
those two equations it follows that, to fulfill the purpose of identifying
distinct configurations having the same value of $E$, we must have
\begin{equation}
3(n_1-n'_1)(1+\delta_1)\neq 2(n'_2-n_2)(1+\delta_2)
\label{e1e2}
\end{equation}
for all possible configurations $n_1,n_2$ and $n'_1,n'_2$ with $n_1\neq n'_1$
or $n_2\neq n'_2$. We use two randomly generated numbers of the order of
$10^{-7}$, viz., $\delta_1=7.72453\times 10^{-7}$ and
$\delta_2=1.47577\times 10^{-7}$, which makes it impossible for the condition in
Eq.~(\ref{e1e2}) to be violated. Note that counting tiles in this slightly
warped manner can be used directly in the nondegenerate case, by letting
\begin{equation}
\hat S_\mathrm{nd}(n_1,n_2)=\mathrm{hist}_{\hat S}(\varepsilon),
\end{equation}
where $\varepsilon$ is the energy level of configuration $n_1,n_2$. It can also
be used in the degenerate case, by coalescing together all energy levels having
the same integral part. If $\mathcal{E}(\bar\varepsilon)$ is the set of all
energy levels sharing the same integral part $\bar\varepsilon$, then coalescing
means letting
\begin{equation}
\hat S_\mathrm{d}(\bar\varepsilon)= \ln\hat K(\bar\varepsilon),
\label{sdginth}
\end{equation}
where
\begin{equation}
\hat K(\bar\varepsilon)=
\sum_{\varepsilon'\in\mathcal{E}(\bar\varepsilon)}
e^{\mathrm{hist}_{\hat S}(\varepsilon')}.
\end{equation}
This quantity is the counterpart, for when analytical treatment is not possible,
of the $K(\varepsilon)$ defined in Eq.~(\ref{Kh}).

\subsection{The $\bm m\bm =\bm 4$ and $\bm m\bm =\bm n$ cases}

In Secs.\ref{ssec:ndg} and~\ref{ssec:dg-4+nxn}, we give results of the
Wang-Landau method for $4\times n$ and $n\times n$ boards. As previously, the
possible values of $E$ in Eq.~(\ref{h}), now considering only their integral
parts, range from a minimum that uses as many tetraminoes as possible, plus as
few dominoes and squares as possible, to a maximum that uses squares
exclusively.

For a $4\times n$ board, this minimum occurs for configuration $n_1=n_2=0,n_4=n$
and equals
\begin{equation}
E_\mathrm{min}^{4\times n}=4(n-1).
\end{equation}
The maximum occurs for configuration $n_1=4n,n_2=n_4=0$ and equals
\begin{equation}
E_\mathrm{max}^{4\times n}=7n-4.
\end{equation}
As for an $n\times n$ board, first let $a=n\bmod 4$. The minimum integral part
of an energy level can be seen to occur for configuration
$n_1=a^2\bmod 2,n_2=\lfloor a^2/2\rfloor,n_4=\lfloor n/4\rfloor(n+a)$, which
yields
\begin{align} 
E_\mathrm{min}^{n\times n}
={}& 2(a^2-2\lfloor a^2/2\rfloor)+3\lfloor a^2/2\rfloor+{} \\
& 5\lfloor n/4\rfloor(n+a)-2n \nonumber\\
={}& 2a^2-\lfloor a^2/2\rfloor+5\lfloor n/4\rfloor(n+a)-2n.
\end{align}
The maximum, in turn, occurs for configuration $n_1=n^2,n_2=n_4=0$, yielding
\begin{equation}
E_\mathrm{max}^{n\times n}=2n(n-1).
\end{equation}

Our computational experiments on $4\times n$ and $n\times n$ boards were planned
so that a board's number of cells would not exceed $256$, as in
Sec.~\ref{sec:spe}. Whenever the number of cells in use happens to be both a
perfect square and a multiple of $16$ (as is $256$), a curious property, using
$n_\mathrm{cells}$ to denote the number of cells, is that
\begin{align}
\frac{3n_\mathrm{cells}}{4}
&= E_\mathrm{max}^{1\times n_\mathrm{cells}}-
E_\mathrm{min}^{1\times n_\mathrm{cells}} \\
&= E_\mathrm{max}^{4\times\frac{n_\mathrm{cells}}{4}}-
E_\mathrm{min}^{4\times\frac{n_\mathrm{cells}}{4}} \\
&= E_\mathrm{max}^{\sqrt{n_\mathrm{cells}}\times\sqrt{n_\mathrm{cells}}}-
E_\mathrm{min}^{\sqrt{n_\mathrm{cells}}\times\sqrt{n_\mathrm{cells}}}.
\end{align}
That is, the size of the energy spectrum is the same in all three cases. Also,
and not surprisingly, the configurations for which the minimum and maximum value
of $E$ are obtained are the same in all three cases:
$n_1=0,n_2=0,n_4=4^{-1}n_\mathrm{cells}$ for the minimum,
$n_1=n_\mathrm{cells},n_2=n_4=0$ for the maximum.

Fixing $n_\mathrm{cells}$ at $256$ reveals further similarities. Not only are
the two configurations of minimum and maximum energy the same in all three
cases, but the set of $4\,225$ configurations of the $1\times 256$ case is the
set of configurations of the $4\times 64$ and $16\times 16$ cases as well. To
see that this is indeed the case, first consider that any configuration of
either the $4\times 64$ or the $16\times 16$ case is also a configuration of the
$1\times 256$ case (simply arrange the tiles in the $1\times 256$ board
arbitrarily). Conversely, given that both $m$ and $n$ are multiples of $4$ in
the $4\times 64$ and $16\times 16$ cases, then any configuration of the
$1\times 256$ case is also a configuration of both the $4\times 64$ and
$16\times 16$ cases. This too is seen to be straightforward, e.g.: first arrange
the $n_4$ tetraminoes in columns of at most $m/4$ tiles each, then fill the
remaining (partially filled or empty) columns with the $n_2$ dominoes and the
$n_1$ squares. 

\begin{figure}[t]
\includegraphics[scale=0.40]{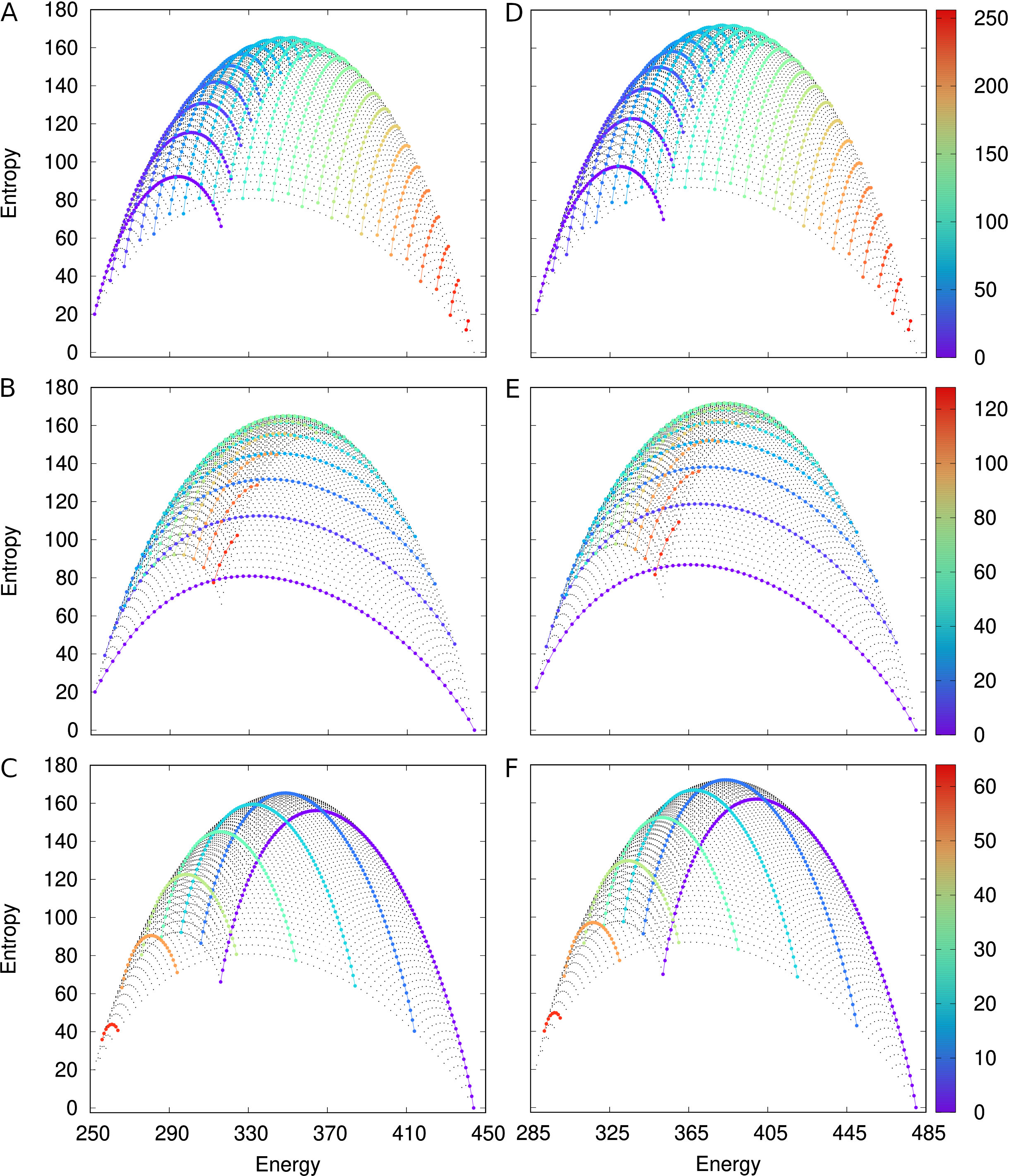}
\caption{As in Fig.~\ref{ndg-1xn}, now for two cases of $m>1$ with
$n_\mathrm{cells}=256$, each configuration represented by its energy $E$ and
entropy $\hat S_\mathrm{nd}$. (A--C) $m=4$, $n=64$,
$E_\mathrm{min}^{4\times 64}=252$, $E_\mathrm{max}^{4\times 64}=444$;
(D--F) $n=m=16$,
$E_\mathrm{min}^{16\times 16}=288$, $E_\mathrm{max}^{16\times 16}=480$.
Color codes refer to multiple-of-$10$ values of $n_1$ (A, D), $n_2$ (B, E), or
$n_4$ (C, F).}
\label{ndg-4+nxn}
\end{figure}

These further similarities between the $1\times 256$, $4\times 64$, and
$16\times 16$ cases are important because they allow us to check the results
output by the Wang-Landau method on the $4\times 64$ and $16\times 16$ boards
against those we already validated analytically for the $1\times 256$ board. The
only differences we expect are significantly higher numbers of tilings, i.e.,
higher entropies and the corresponding adjustments in temperature. All else is
expected to remain unaltered.

\subsection{The nondegenerate case}
\label{ssec:ndg}

Our results from the Wang-Landau method on the $4\times 64$ and $16\times 16$
boards, when degeneracy is disregarded so that partitions of the configuration
space can be observed, are summarized in the panels of Fig.~\ref{ndg-4+nxn}.
Each panel has exactly $4\,225$ background dots, one for each configuration,
indicating that the MCMC walker reached all of them. The likeness of the set of
panels corresponding to the $4\times 64$ board (A--C), or of those corresponding
to the $16\times 16$ board (D--F), to Fig.~\ref{ndg-1xn} cannot be missed. In
fact, as expected, only the entropy values help distinguish one case from the
other two. While for the $1\times 256$ board we have $S_\mathrm{nd}<140$, for
the $4\times 64$ and $16\times 16$ boards we have $\hat S_\mathrm{nd}<166$ and
$\hat S_\mathrm{nd}<173$, respectively.

\subsection{The degenerate case}
\label{ssec:dg-4+nxn}

\begin{figure}[t]
\includegraphics[scale=0.40]{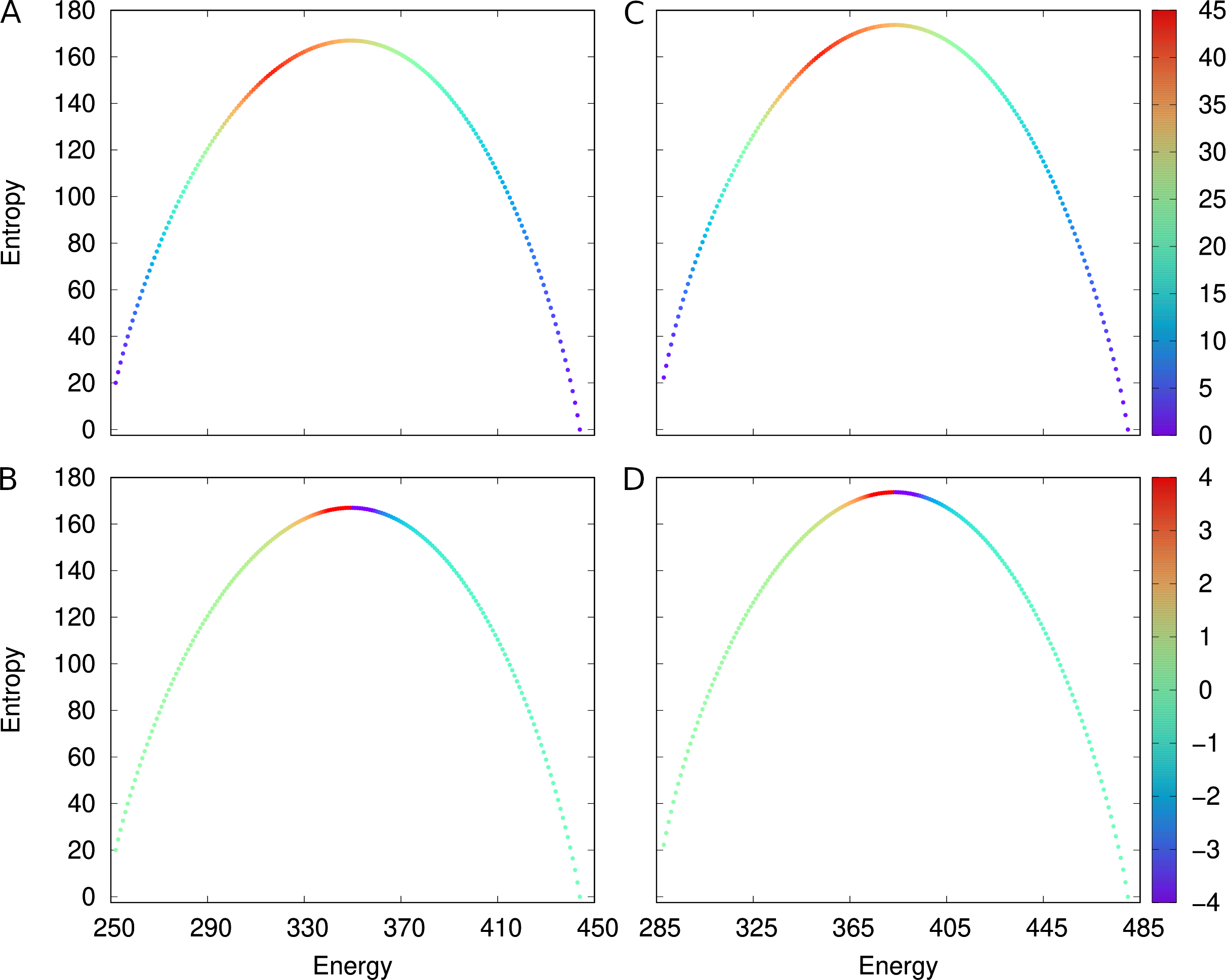}
\caption{As in Fig.~\ref{deg-1xn}, now showing entropy $\hat S_\mathrm{d}$ as a
function of energy $E$ for two cases of $m>1$ with $n_\mathrm{cells}=256$.
(A, B) $m=4$, $n=64$,
$E_\mathrm{min}^{4\times 64}=252$, $E_\mathrm{max}^{4\times 64}=444$;
(C, D) $m=n=16$,
$E_\mathrm{min}^{16\times 16}=288$, $E_\mathrm{max}^{16\times 16}=480$.
Color codes refer to numbers of configurations (A, C) or to temperature
$\hat T_\mathrm{d}$ (B, D).}
\label{deg-4+nxn}
\end{figure}

When degeneracy is taken into account and all configurations for the same energy
level are coalesced together, using the Wang-Landau method on the $4\times 64$
and $16\times 16$ boards yields the results shown in Fig.~\ref{deg-4+nxn}. This
figure has two panels for the $4\times 64$ board (A, B) and two for the
$16\times 16$ board (C, D). Two of the panels highlight the number of
configurations contributing to each entropy value (A, C) and two others
highlight the corresponding temperature (B, D). Temperature is now estimated
from entropy differences, based on substituting the $\hat S_\mathrm{d}$ of
Eq.~(\ref{sdginth}) for $S_\mathrm{d}$ in Eq.~(\ref{tdgm1d}). This results in
\begin{equation}
\hat T_\mathrm{d}^{-1}(\bar\varepsilon)\approx
\frac
{\hat S_\mathrm{d}(\bar\varepsilon+1)-\hat S_\mathrm{d}(\bar\varepsilon-1)}
{2}
\end{equation}
for the combined energy levels of integral part $\bar\varepsilon$.

Once again, the resemblance of all four plots to the corresponding ones in
Fig.~\ref{deg-1xn} is hard to miss. Differences do exist, however, the clearest
one relating to the maximum entropy levels in each case, as noted in
Sec.~\ref{ssec:ndg}. The other difference has to do with the temperature
estimates, which in all three cases are color-coded in the lower panels, but are
nevertheless hard to discern visually.

\subsection{Further remarks on computational tractability}

\begin{figure}[t]
\includegraphics[scale=0.68]{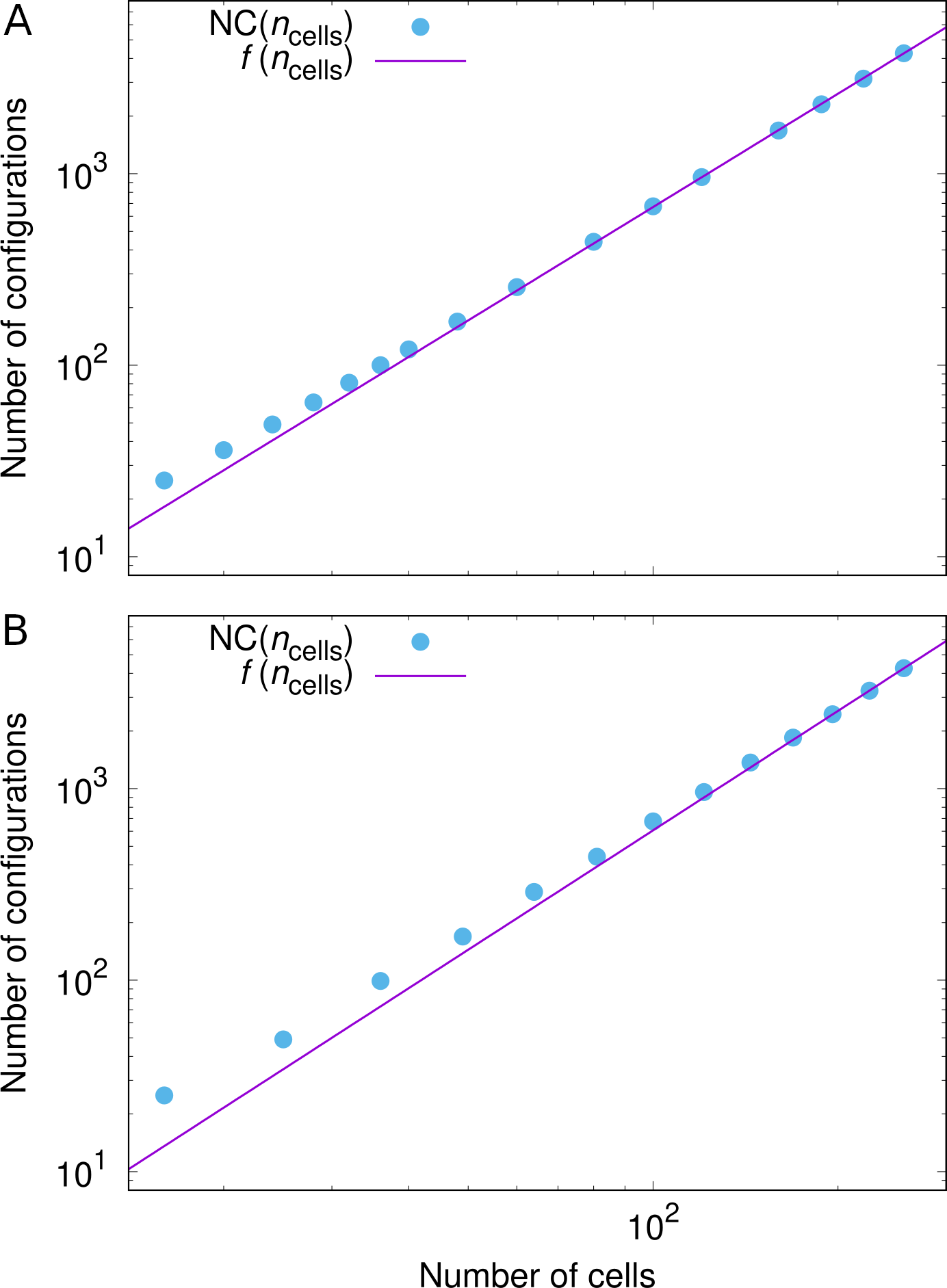}
\caption{As in Fig.~\ref{nconfigs-1xn}, now as the function
$\mathrm{NC}(n_\mathrm{cells})$ for two cases of $m>1$ with
$n_\mathrm{cells}\le 256$. (A) $m=4$, $n=4^{-1}n_\mathrm{cells}$; (B)
$m=n=\sqrt{n_\mathrm{cells}}$. The power law that fits the points asymptotically
is $f(n_\mathrm{cells})=12.79^{-1}n_\mathrm{cells}^{1.96648}$ (A)
or $f(n_\mathrm{cells})=23^{-1}n_\mathrm{cells}^{2.07212}$ (B).}
\label{nconfigs-4+nxn}
\end{figure}

We finalize Sec.~\ref{sec:gen} by returning to the theme of computational
tractability raised in Sec.~\ref{ssec:ct} in the context of the $m=1$ special
case, for which analytical treatment is possible. The concern in that case was
centered on the total number of configurations. These had to be enumerated to
exhaustion for use in the analyses, a process that we found out is severely
limited as the number of cells in the board grows. The total number of tilings
$k_{1n}$ for each configuration, from which entropy is calculated, was in that
case reason for no concern in terms of computational tractability, since the
availability of a closed-form expression for $k_{1n}$ was itself the one key
factor enabling analytical treatment.

\begin{figure}[t]
\includegraphics[scale=0.68]{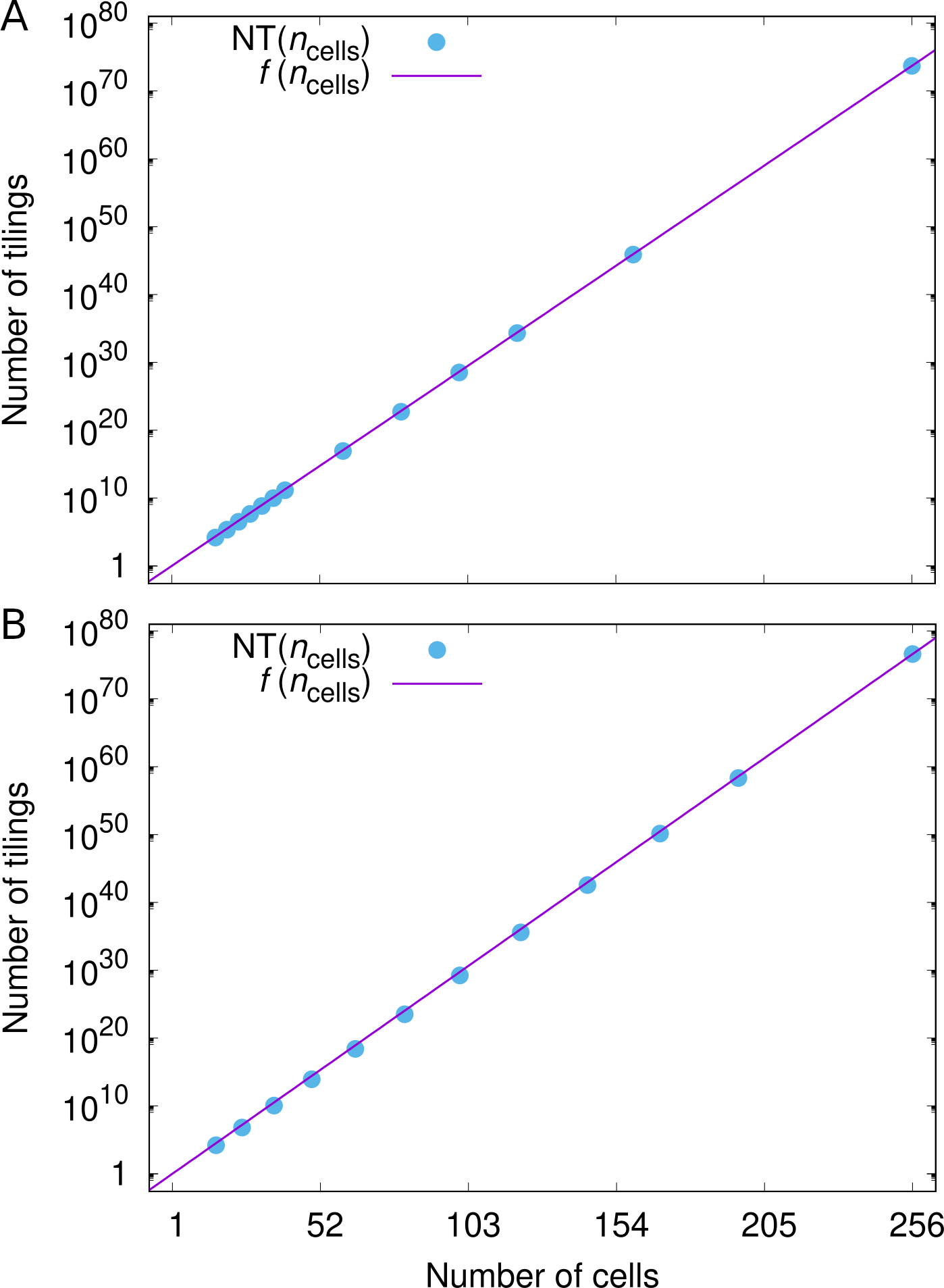}
\caption{As in Fig.~\ref{ntilings-1xn}, now as the function
$\mathrm{NT}(n_\mathrm{cells})=\sum_{\bar\varepsilon}\hat K(\bar\varepsilon)$
for two cases of $m>1$ with $n_\mathrm{cells}\le 256$. (A) $m=4$,
$n=4^{-1}n_\mathrm{cells}$; (B) $m=n=\sqrt{n_\mathrm{cells}}$. The exponential
that fits the points is
$f(n_\mathrm{cells})=e^{0.665527(n_\mathrm{cells}-1)}$ (A) or
$f(n_\mathrm{cells})=e^{0.691678(n_\mathrm{cells}-1)}$ (B).}
\label{ntilings-4+nxn}
\end{figure}

In the context of the $m>1$ cases we have been handling in Sec.~\ref{sec:gen}, 
the main concern related to computational tractability is the growth of the
tiling space through which the MCMC walker navigates to estimate entropy. This
space grows unimaginably quickly with both the total number of configurations
for the different energy levels and especially the total number of tilings
admitted by those configurations. A larger tiling space requires more steps for
the walker to be able to roam sufficiently far and wide for convergence to
occur. Figures~\ref{nconfigs-4+nxn} and~\ref{ntilings-4+nxn} illustrate the
growth trends of the two quantities.

Figure~\ref{nconfigs-4+nxn} is about the growth of the total number of
configurations as $n_\mathrm{cells}$ grows. Panel A is for
$4\times\frac{n_\mathrm{cells}}{4}$ boards, panel B for
$\sqrt{n_\mathrm{cells}}\times\sqrt{n_\mathrm{cells}}$ boards. For each of these
two cases we found the total number of configurations to asymptotically follow a
power law similar to the quadratic one found for the $1\times n_\mathrm{cells}$
but still different from it. Figure~\ref{ntilings-4+nxn}, in turn, is about the
growth of the total number of tilings as $n_\mathrm{cells}$ grows, with panels
arranged analogously to Fig.~\ref{nconfigs-4+nxn}. Exponentials in
$n_\mathrm{cells}$ are still found, but now reaching significantly higher
numbers. For $n_\mathrm{cells}=256$, a total number of tilings of the order of
$10^{73}$ was found in the $4\times\frac{n_\mathrm{cells}}{4}$ case, of the
order of $10^{76}$ in the $\sqrt{n_\mathrm{cells}}\times\sqrt{n_\mathrm{cells}}$
case. These are to be compared with the total number of tilings in the
$1\times n_\mathrm{cells}$ case, which is of the order of $10^{63}$.

\section{Conclusion}
\label{sec:con}

The scientific interest in tilings of the plane, and sometimes of
higher-dimensional regions as well, has a history that spans several centuries.
In the last six to seven decades, however, it gained new momentum motivated by
the realization in the early 20th century that some decision problems could be
neither easy nor hard to solve, but simply undecidable. Though it soon turned
out that undecidability was not an issue, interest did not wane. Instead, due to
efforts by both mathematicians working on discrete systems and applied
physicists in fields related to atomic- and molecular-scale phenomena, progress
with important milestones was maintained. By and large, however, it seems fair
to say that most key theoretical advancements, even those that inspired
real-world applications, came from considering tilings of the whole, unbounded
plane.

Studying tilings of bounded regions of the plane requires handling the
combinatorics of finite discrete structures, a field in which exact closed-form
expressions for counting quantities of interest are hard to come by. In this
study, we considered rectangular regions of the plane and how to tile them with
rectangular tiles. We concentrated on the tile-type set containing the square,
the domino, and the straight tetramino. Simple though this system may seem, in
the general case of an $m\times n$ board not even the number of configurations
(how many squares, dominoes, tetraminoes) that are feasible can be counted
exactly, not even indirectly via recurrence relations. Our approach has been to
regard the system from the standpoint of their statistical properties and
consider its configurations, states (tilings), energy, entropy, and temperature
in such a way as to illuminate some of its inner workings. We followed two
parallel tracks, one for the $m=1$ case, the other for more general, $m>1$
cases. Given our choice of a tile contact-based energy function, the $m=1$ case
is fully tractable analytically. The other track relied on the Wang-Landau
method for state-density estimation in the $m>1$ cases, and on the subsequent
calculation of approximate entropies and temperatures. By alternately
disregarding and taking into account the issue of energy-level degeneracy, we
were able to demonstrate how to partition the configuration space and thereby
highlight how the system's configurations relate to one another, as well as to
entropy and temperature.

A lot of room is left for methodological improvements. In particular, given the
computationally intensive character of the Wang-Landau method, some of the
techniques already developed for exploring the state space in parallel
\cite{vlwl14} should be considered. Additionally, our choice of energy function
has been about the simplest imaginable. Considering the next level of
sophistication, by adopting an energy function based on tile areas instead of
simply perimeters, is bound to bring the entire approach closer to some of the
applications that might benefit from it. Some of these applications are in areas
such as modeling biological tissues \cite{fraej07,bhws17} and their properties
\cite{csmbg20,pkbmqtpmkgnsbrkthsiwthwmbdf15}, and developing bioinspired
metamaterials \cite{p21}.

\begin{acknowledgments}
This work is part of the INCT-Física Nuclear e Aplicações project,
No.~464898/2014-5. We acknowledge partial support from Conselho Nacional de
Desenvolvimento Científico e Tecnológico (CNPq), Coordenação de Aperfeiçooamento
de Pessoal de Nível Superior (CAPES), and a BBP grant from Fundação Carlos
Chagas Filho de Amparo à Pesquisa do Estado do Rio de Janeiro (FAPERJ), as well
as support from Agencia Nacional de Investigación e Innovación (ANII) and
Programa de Desarrollo de las Ciencias Básicas (PEDEClBA). We thank Núcleo
Avançado de Computação de Alto Desempenho (NACAD), Instituto Alberto Luiz
Coimbra de Pós-Graduação e Pesquisa em Engenharia (COPPE), Universidade Federal
do Rio de Janeiro (UFRJ), for the use of supercomputer Lobo Carneiro, where most
of the calculations were carried out.
\end{acknowledgments}

\bibliography{tilings}
\bibliographystyle{apsrev4-2}

\end{document}

%% file: table1.tex
\renewcommand{\arraystretch}{1.5}
\setlength{\tabcolsep}{10pt}
\sisetup{table-number-alignment=center,
         group-minimum-digits=4}
\begin{tabular}{S[table-format=2]
                S[table-format=2.2]
                S[table-format=2]
                S[table-format=2]
                S[table-format=1]
                S[table-format=5]
                c}
\hline
{$E$} & {$S_\mathrm{nd}$} & {$n_1$} & {$n_2$} & {$n_4$} & {$k_{1n}^{n_4=1}$}
& One of the $k_{1n}^{n_4=1}$ tilings\\
\hline
11  &   4.88  &   1  &  10  &  1  &    132
& \begin{minipage}{.45\textwidth}
  \includegraphics[width=\linewidth]{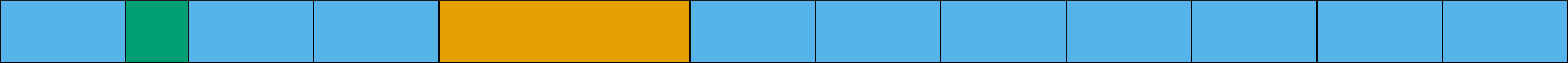}
  \end{minipage} \\
12  &   7.95  &   3  &   9  &  1  &   2860
& \begin{minipage}{.45\textwidth}
  \includegraphics[width=\linewidth]{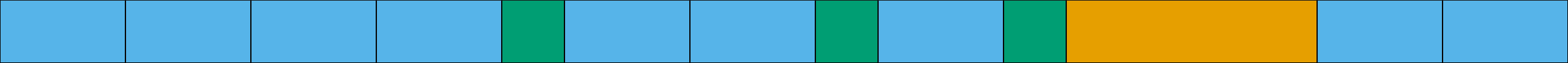}
  \end{minipage} \\
13  &   9.79  &   5  &   8  &  1  &  18018
& \begin{minipage}{.45\textwidth}
  \includegraphics[width=\linewidth]{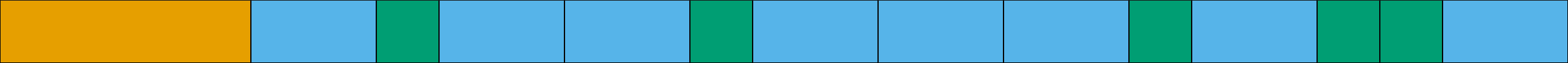}
  \end{minipage} \\
14  &  10.84  &   7  &   7  &  1  &  51480
& \begin{minipage}{.45\textwidth}
  \includegraphics[width=\linewidth]{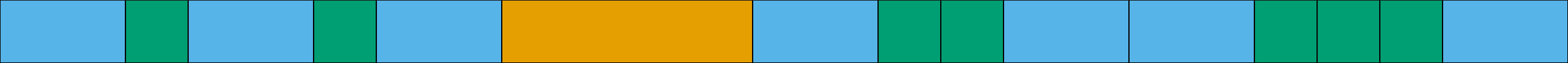}
  \end{minipage} \\
15  &  11.29  &   9  &   6  &  1  &  80080
& \begin{minipage}{.45\textwidth}
  \includegraphics[width=\linewidth]{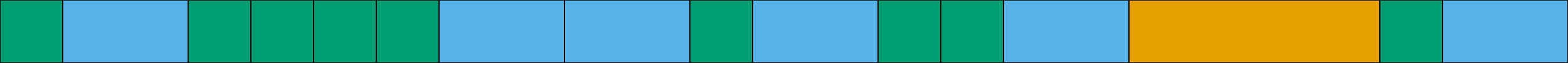}
  \end{minipage} \\
16  &  11.21  &  11  &   5  &  1  &  74256
& \begin{minipage}{.45\textwidth}
  \includegraphics[width=\linewidth]{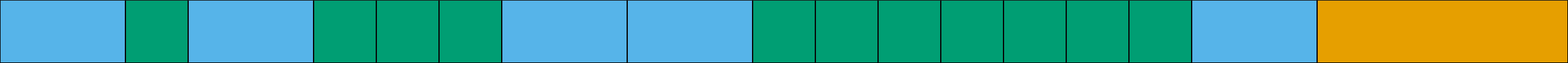}
  \end{minipage} \\
17  &  10.66  &  13  &   4  &  1  &  42840
& \begin{minipage}{.45\textwidth}
  \includegraphics[width=\linewidth]{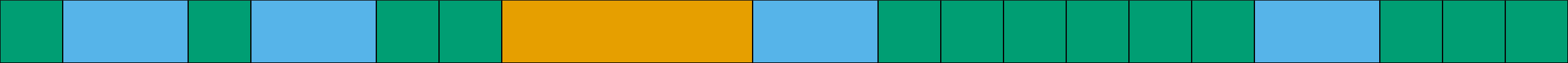}
  \end{minipage} \\
18  &   9.64  &  15  &   3  &  1  &  15504
& \begin{minipage}{.45\textwidth}
  \includegraphics[width=\linewidth]{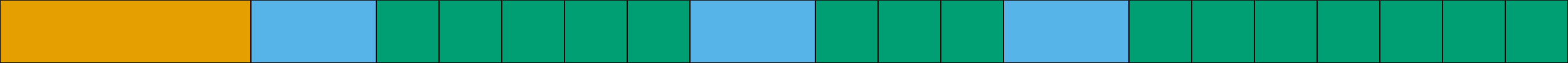}
  \end{minipage} \\
19  &   8.13  &  17  &   2  &  1  &   3420
& \begin{minipage}{.45\textwidth}
  \includegraphics[width=\linewidth]{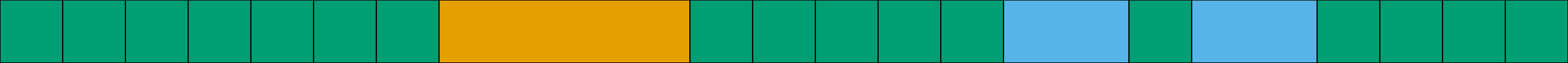}
  \end{minipage} \\
20  &   6.04  &  19  &   1  &  1  &    420
& \begin{minipage}{.45\textwidth}
  \includegraphics[width=\linewidth]{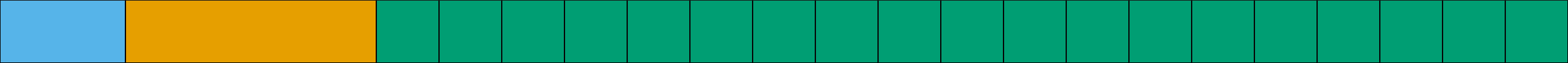}
  \end{minipage} \\
21  &   3.09  &  21  &   0  &  1  &     22
& \begin{minipage}{.45\textwidth}
  \includegraphics[width=\linewidth]{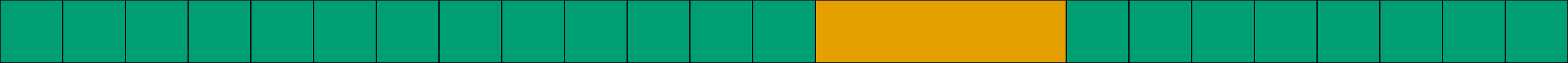}
  \end{minipage} \\
\hline
\end{tabular}